\documentclass[fleqn,twoside]{article}
\usepackage{espcrc2}

\usepackage{graphicx}
\usepackage[figuresright]{rotating}

\newcommand{\AmS}{{\protect\the\textfont2
  A\kern-.1667em\lower.5ex\hbox{M}\kern-.125emS}}

\title{Application of Maximum Entropy Method to Dynamical Fermions}

\author{Jonathan Clowser$^{\rm a}$
         and Costas Strouthos\address{Department of Physics,
        University of Wales Swansea, \\
        Singleton Park, Swansea SA2 8PP, United Kingdom}
        }

\begin{document}

\begin{abstract}
The Maximum Entropy Method is applied to dynamical fermion simulations of the (2+1)-dimensional
Nambu$-$Jona-Lasinio model.
This model is particularly interesting because at $T\!=\!0$ it has a broken phase
with a rich spectrum of mesonic bound states and a symmetric phase where there are resonances,
and hence the simple pole assumption of traditional fitting procedures breaks down.
We present results extracted from simulations on large lattices for the spectral functions of the
elementary fermion, the pion, the sigma, the massive
pseudoscalar meson and the symmetric phase resonances.
\vspace{1pc}
\end{abstract}

\maketitle

\section{Introduction}
The QCD spectral function $\rho(\omega)$ carries information not just about
particle masses but also about resonance widths.
Recently, the first serious attempts
to reconstruct spectral functions, defined via
\begin{equation}
G(t) = \int_{0}^{\infty} K(t, \omega) \rho(\omega) d\omega,
\end{equation}
where the kernel $K$ is a free particle correlator with mass
$\omega$, from lattice data have been made. The inverse problem is
numerically ill-posed, since $\rho(\omega)$ is a continuous
function whereas the lattice propagator $G(t)$ is only available
for a discrete and finite set of points. The procedure adopted has
been to assume an ansatz for $\rho(\omega)$ based on QCD sum rules
\cite{qcdsum}. A more general solution is to apply maximum entropy
method (MEM) techniques \cite{mem}, which extract $\rho(\omega)$
subject to some reasonable requirements that it be smooth and
stable with respect to small variations in the input data $G(t)$,
but otherwise make no a priori assumptions about its form. The
theoretical basis of MEM is Bayes' theorem in probability theory.
Details about the technique can be found in \cite{mem}. A
sophisticated technique of analyzing data such as MEM is also
needed to study the complexity of the mesonic spectral functions
in thermal QCD below $T_c$ and at the range of intermediate
temperatures above $T_c$ which is dominated by strongly
interacting quarks in the mesonic channels.

To refine these new techniques we started an investigation of spectroscopy in general,
and the MEM method in particular, in the $(2\!+\!1)$-dimensional Nambu$-$Jona-Lasinio (NJL)
model. The continuum Euclidean spacetime Lagrangian is:
\begin{eqnarray*}
{\cal L}= \bar{\Psi}_i(\partial\hskip -.5em /  + m_0 + \sigma + i \gamma_5 \pi)\Psi_i
+ \frac{N_f}{2 g^{2}} (\sigma^{2}+ \pi^2).
\end{eqnarray*}
The fields $\Psi_i$ and $\bar{\Psi}_i$ are four-component spinors and the index
$i$ runs over $N_f$ fermion flavours.
The model has an interacting continuum limit for
a critical value of the coupling
$1/g_c^2$, which has a numerical value $\approx 1.0/a$ in the large-$N_f$ limit
if a lattice regularization is employed \cite{hands93}.
At $T\!=\!0$ and for $m_0\!=\!0$ the model has a $U(1)$ chiral symmetry which is broken
spontaneously for sufficiently
strong coupling $g^2 \! > \! g_{c}^2 $ and the pion field $\pi$ is the associated
Goldstone boson \cite{rosen91}.
Despite its simplicity the NJL model has an interesting
particle spectrum. The quark propagator is gauge invariant and easily measured,
enabling the fermion mass to be monitored as chiral symmetry is restored. The
$\sigma$ and $\pi$ mesons are represented by bosonic auxiliary fields in the
code, so the correlation functions in these channels, including the disconnected
diagrams so expensive to calculate in QCD, can be measured with high statistics
relatively cheaply.
The Goldstone mechanism in the NJL model is fundamentally different from that in
QCD. In the NJL model the disconnected diagrams are responsible for making
the pion light \cite{hands99}. The connected  contribution corresponds to a pseudoscalar state
with a mass which is twice the dynamical fermion mass in the large-$N_f$ limit.
Another interesting aspect of the model is that in the large-$N_f$ limit
the correlator of the $\sigma$ meson has a complicated
spatial dependence due to a strongly interacting  fermion-antifermion continuum
for $k\!>\!M_{\sigma}\!=\!2\langle \sigma \rangle$ \cite{rosen91,hands93}, i.e. it has  branch cuts
in the complex $k$-plane in the ranges $[iM_{\sigma}, i\infty)$ and
$(-i\infty, -iM_{\sigma}]$, instead of a discrete series of poles,
corresponding to a set of bound states in the channel in question,
which is the case of QCD.
In our simulations we check whether next-to-leading order effects
make the $\sigma$ meson stable by separating a bound state pole from the
two-fermion threshold.
The NJL model has also an interesting symmetric phase. In this phase the large-$N_f$ meson
correlator does not interpolate as a stable massive particle state, but rather
an unstable resonance and the resonance width serves as an inverse correlation
length \cite{rosen91,hands93}.

\section{Lattice simulations and MEM analysis}
The model was discretized by using the staggered fermion formulation and the simulations
were performed by using the standard Hybrid Monte Carlo algorithm.
The lattice action and details concerning the algorithm can be found in \cite{hands93,hands99,hands95}.
We performed simulations with both large $N_f\!=\!36$ and relatively small $N_f\!=\!4$.
This enables us first to compare the MEM results with the analytical large-$N_f$ predictions
and then to use the method to probe
next-to-leading order corrections.
The propagators for the auxiliary fields $\pi$ and $\sigma$ are given by:
\begin{eqnarray}
C_{\pi}(t)=\sum_{\mathbf{x},t^{\prime}} \langle \pi(\mathbf{x},t^{\prime})
\pi(\mathbf{x},t^{\prime}+t) \rangle, \\
C_{\sigma}(t)=\sum_{\mathbf{x},t^{\prime}}[\langle \sigma(\mathbf{x},t^{\prime})
\sigma(\mathbf{x},t^{\prime}+t)\rangle - \langle \sigma \rangle^2 ].
\end{eqnarray}
In order to minimize contamination from unwanted excited states
the massive PS meson correlator was measured by using wall sources.
%

\begin{figure}[htb]
\centering
\includegraphics[scale=0.22]{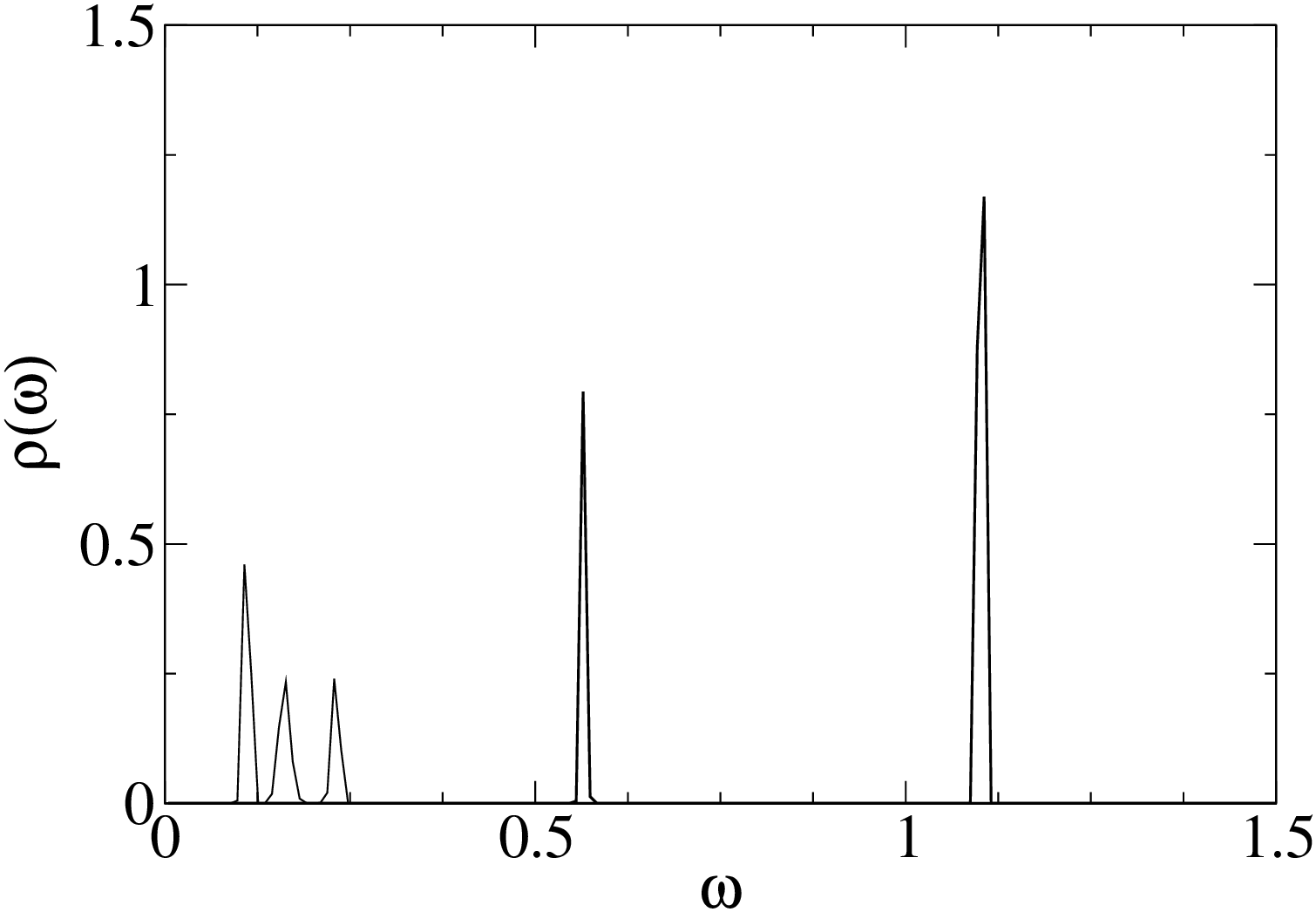}
\vspace{-1cm}
\caption{\small Broken phase spectral functions, peaks from left to right are for the
         $\pi$ ($m_0=0.005,0.01,0.02$), f and PS (both $m_0=0.01$).}
\label{fig:broken}
\end{figure}

Figure~\ref{fig:broken} shows the spectral functions output from MEM for the
$\pi$, fermion (f) and massive PS meson in the broken phase of $N_f\!=\!4$ on a lattice
of size $32^2\!\times \! 48$ and coupling $\beta \equiv 1/g^2 \!=\! 0.55$. We generated
approximately $30,000$ configurations for each parameter set.
\vspace{-0.3cm}
\vspace{-1.0cm}
\begin{figure}[htb]
\centering
\includegraphics[scale=0.22]{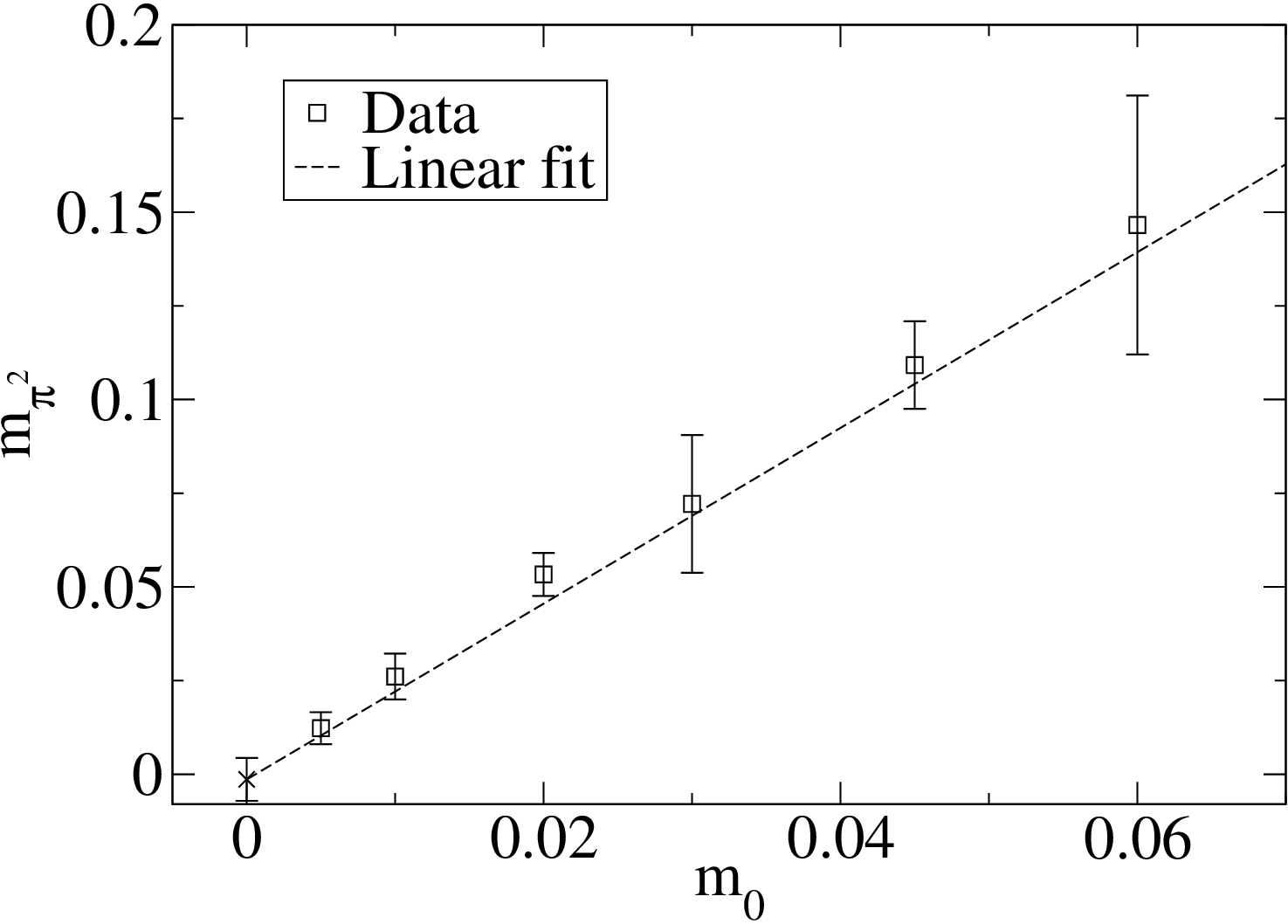}
\vspace{-1cm}
\caption{\small PCAC scaling relation for $\beta=0.55$ data sets.}
\label{fig:pcac}
\end{figure}
\vspace{-0.7cm}

These particles are all expected to be poles.
The masses extracted from the MEM analysis are in very good agreement
with standard single-exponential fits.
(e.g. for the $\pi$ in $N_f \! = \! 4$, $\beta \! = \! 0.55$,
$m_0 \! = \! 0.01$: 1-exp=0.168(6),
MEM=0.162(6) and for $m_0=0.02$: 1-exp=0.236(6), MEM=0.231(6)). In order to check that the
width of the peaks seen in Figure~\ref{fig:broken} are statistical, an extrapolation
to an infinite number of configurations will be performed.
The binding energy, $\epsilon_b$, for the massive PS meson is calculated from $2M_f-M_{PS}$.
For $\beta=0.55$ $m_0=0.01$:
$\epsilon_b=0.027$ for $N_f=4$ and $\epsilon_b=0.006$ for $N_f=36$.
The result is in agreement with the expectation that the meson is more tightly
bound for smaller $N_f$.
Figure~\ref{fig:pcac} shows that the
pion mass obeys the PCAC scaling relationship since the linear fit intercepts the y-axis at
-0.001(6).
%

\begin{figure}[htb]
\centering
\includegraphics[scale=0.22]{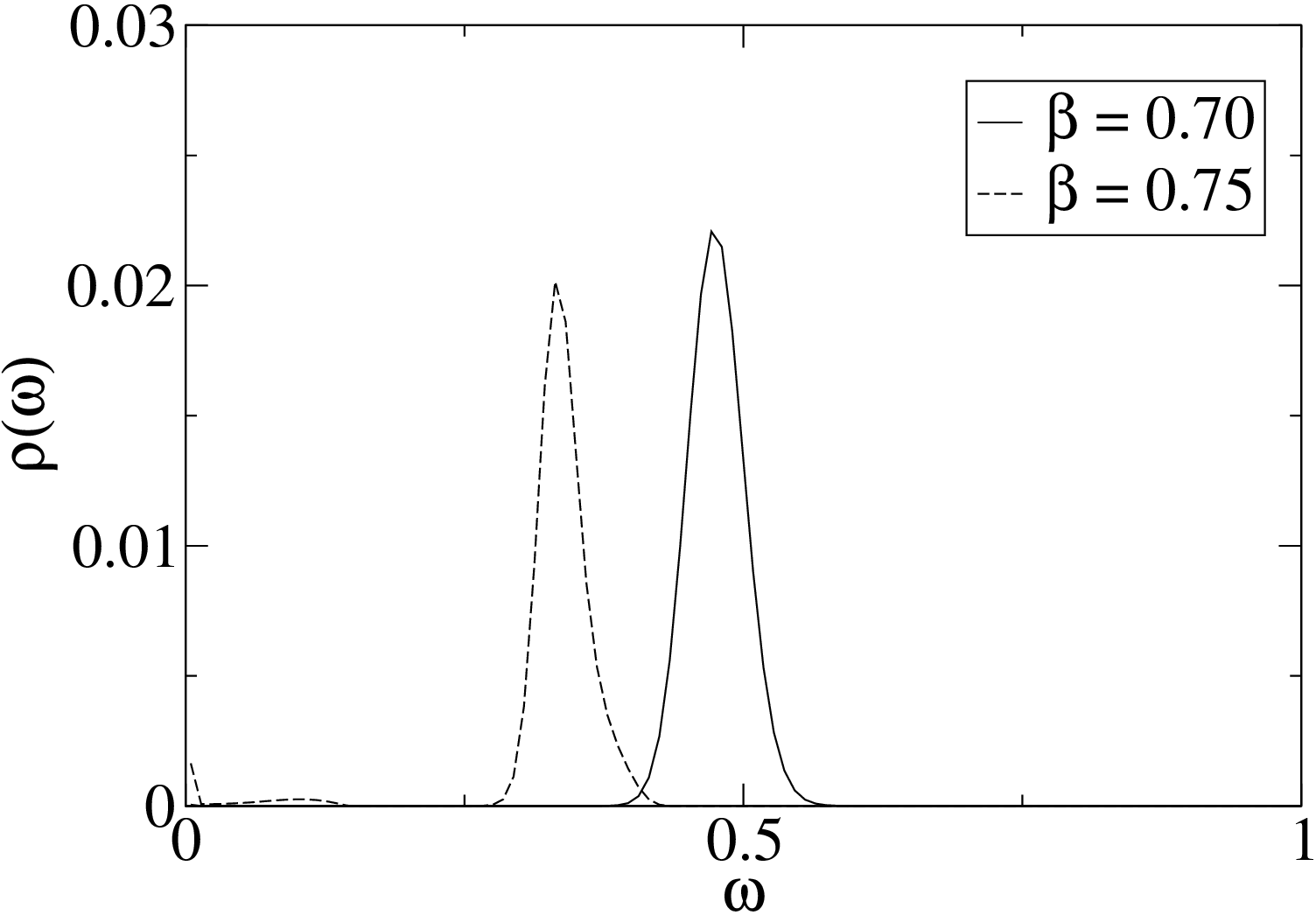}
\vspace{-1cm}
\caption{\small Sigma meson in $Z_2$ model, $N_f=4$.}
\label{fig:sigma}
\end{figure}

For the $\sigma$ meson we generated very high statistics
($\sim \!{\cal O}(10^5)$) on a $24^3$ lattice using the simpler $Z_2$-symmetric
(rather than $U(1)$-symmetric) model. It is not clear yet whether the widths of the spectral functions
shown in Figure ~\ref{fig:sigma} are due to the noise in our data set or whether they have a physical
significance. However, the positions of the peaks imply that the sigma is a bound state, because e.g.,
for $\beta=0.70$, $\omega_{peak}=0.475(16)$ is significantly smaller than $2M_f\!=\!0.534(1)$. This is
another next-to-leading order effect that we detected in our MEM analysis.

\vspace{-0.3cm}
\vspace{-0.7cm}

\begin{figure}[t!]
\centering
\includegraphics[scale=0.22]{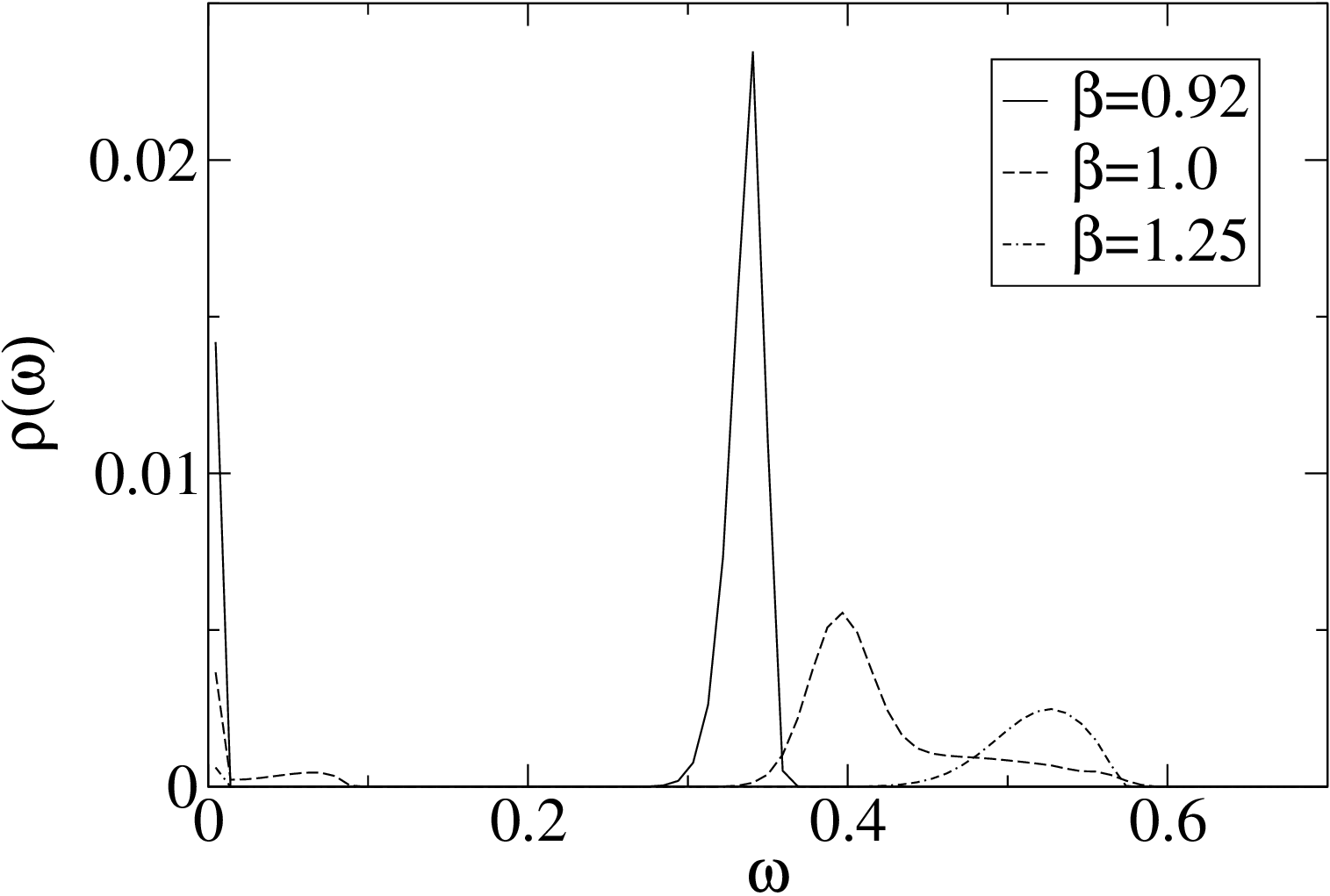}
\vspace{-1cm}
\caption{\small Symmetric phase scalar for $N_f=4$.}
\label{fig:symmetric4}
\end{figure}

In Figure~\ref{fig:symmetric4} the results obtained for the
symmetric phase of the $Z_2$ model ($N_f=4$ on a $32^2\times 48$ lattice)
are presented. The critical coupling is $\beta_c=0.84(1)$
\cite{strouthos}
and the simulations were performed with $\beta=0.92, 1.0, 1.25$.
The expected features (in the large-$N_f$ limit)
are a broad resonance whose central position and width increase with $\beta$.
It is not clear yet whether the peaks near the origin have a physical significance or
whether they are finite $L_t$ effects. The correlators in the symmetric phase decay very
slowly and remain significantly non-zero at large $t$. This behaviour is
consistent with the power-law form prediction of the large-$N_f$ calculation \cite{hands93}.
We are currently performing new simulations on lattices with both smaller and bigger $L_t$
in order to gain a better understanding of $\rho(\omega)$ near the origin.

\section{Summary and Outlook}
The preliminary results presented in this report show that MEM is a reliable
method to understand the details of the particle spectra of NJL and other
models.
We are currently increasing our statistics and also intend to calculate errors on
the spectral features in the MEM results.
Furthermore, we are also planning to extend our investigation to understand the spectral properties
of the mesons at high temperature and density.

\section*{Acknowledgements}
This project is being done in collaboration with Chris Allton, Simon Hands
and John Kogut. Costas Strouthos is supported by a Leverhulme Trust grant.

\end{document}